\documentclass[reprint,article,superscriptaddress,msmath,amssymb,aps]{revtex4-1}
\usepackage{graphicx}
\usepackage{dcolumn}
\usepackage{bm}
\usepackage{color}
\usepackage[]{natbib}
\usepackage{amsmath,amssymb,amsfonts}
\usepackage{float}

\begin{document}

\preprint{APS/123-QED}

\title{Primary thermometry of a single reservoir \\ using cyclic electron tunneling in a CMOS transistor}

\author{Imtiaz Ahmed}
\affiliation{Cavendish Laboratory, University of Cambridge, J. J. Thomson Ave., Cambridge, CB3 0HE, United Kingdom}
\author{Anasua Chatterjee}
\affiliation{London Centre for Nanotechnology, University College London, London, WC1H 0AH, United Kingdom}
\affiliation{Center for Quantum Devices, Niels Bohr Institute, University of Copenhagen, 2100 Copenhagen, Denmark}
\author{Sylvain Barraud}
\affiliation{CEA/LETI-MINATEC, CEA-Grenoble, 38000 Grenoble, France}
\author{John~J.~L.~Morton }
\affiliation{London Centre for Nanotechnology, University College London, London, WC1H 0AH, United Kingdom}
\affiliation{Department of Electronic \& Electrical Engineering, University College London, London WC1E 7JE, United Kingdom}
\author{James~A.~Haigh}
\affiliation{Hitachi Cambridge Laboratory, J. J. Thomson Ave., Cambridge, CB3 0HE, United Kingdom}
\author{M. Fernando Gonzalez-Zalba}
\affiliation{Hitachi Cambridge Laboratory, J. J. Thomson Ave., Cambridge, CB3 0HE, United Kingdom}

\begin{abstract}

Temperature is a fundamental parameter in the study of physical phenomena. At the nanoscale, local temperature differences can be harnessed to design novel thermal nanoelectronic devices or test quantum thermodynamical concepts. Determining temperature locally is hence of particular relevance. Here, we present a primary electron thermometer that allows probing the local temperature of a single electron reservoir in single-electron devices. The thermometer is based on cyclic electron tunneling between a system with discrete energy levels and a single electron reservoir. When driven at a finite rate, close to a charge degeneracy point, the system behaves like a variable capacitor whose magnitude and line-shape varies with temperature. In this experiment, we demonstrate this type of thermometer using a quantum dot in a CMOS nanowire transistor. We drive cyclic electron tunneling by embedding the device in a radio-frequency resonator which in turn allows us to read the thermometer dispersively. We find that the full width at half maximum of the resonator phase response depends linearly with temperature via well known physical law by using the ratio $k_\text{B}/e$ between the Boltzmann constant and the electron charge. Overall, the thermometer shows potential for local probing of fast heat dynamics in nanoelectronic devices and for seamless integration with silicon-based quantum circuits.

\end{abstract}

\maketitle

  
\section{\label{sec:level2}Introduction}

An essential element in low temperature experimental physics is the thermometer\cite{Engert2016}. Sensors that link temperature to another physical quantity in an accurate, fast, stable and compact manner are desired. If the link is done via a well known physical law, the sensor is a called a primary thermometer because it removes the need of calibration to a second thermometer. 
 
Several primary thermometers have been developed for low temperature applications. A common technique is based on the Johnson-Nyquist noise of a resistor~\cite{white1984systematic} which can be used in combination with superconducting quantum interference devices to perform current-sensing noise thermometry (CSNT)~\cite{Shibahara2016}. Shot-noise thermometry (SNT)~\cite{spietz2003primary,spietz2006shot,Iftikhar2016} uses the temperature-dependent voltage scaling of the noise power of a biased tunnel junction. Coulomb blockade thermometry (CBT) makes use of charging effects in two-terminal devices with multiple tunnel junctions~\cite{pekola1994thermometry,bradley2016nanoelectronic,hahtela2016traceable}. Thermometry using counting statistics via single-electron devices is also possible~\cite{maradan2014gaas,mavalankar2013non,gasparinetti2012nongalvanic,schmidt2003nanoscale, rossi2012electron}. However, in all these cases, the sensors require a continuous flow of electrons from source to drain in two terminal devices which, for particular experiments such as in single-molecule junction and single-nanoparticle devices, might not be possible or even desirable \cite{aradhya2013single, frake2015radio}. 


Moreover, recent advances in device nanoengineering have led to a focused interest in using concepts from quantum thermodynamics~\cite{Giazotto2012,Pekola2015,Partanen2016,Dutta2017,Sivre2017} to improve the efficiency of technologies such as the thermal diode~\cite{Martinez-Perez2015,Marcos-vicioso2018} or thermal energy harvesters~\cite{Thierschmann2015}. In these nanoelectronic devices, determining the local temperature in different reservoirs of the device is of particular relevance but challenging from an experimental perspective. 

Here, we demonstrate a novel type of primary thermometer that uses cyclic electron tunneling to measure the temperature of a single electron reservoir without the need of electrical transport. The tunneling occurs between a system with a zero-dimensional (0D) density of states (DOS) --in this case a quantum dot (QD) -- and a single electron reservoir of unknown temperature. Our thermometer relates temperature and capacitance changes with a well known physical law by using the ratio $k_\text{B}/e$ between the Boltzmann constant and the electron charge. The thermometer is driven and read out by an electrical resonator at radio-frequencies. In this proof-of-principle experiment, we perform primary thermometry down to 1~K but show that the operational temperature range of the sensor can be extended \lq\lq in-situ\rq\rq\ using electrostatic fields. Our experimental results follow our theoretical predictions of the temperature-dependent capacitance of the system. The thermometer is implemented in a complementary-metal-oxide-semiconductor (CMOS) transistor which makes it suitable for large-scale manufacturing and seamless integration with silicon-based quantum circuits, a promising platform for the implementation of a scalable quantum computer~\cite{Vandersypen2017,Veldhorst2017,Li2017}.

\section{\label{sec:level2} Theory  }

\begin{figure}
\begin{center}
\includegraphics[scale=1]{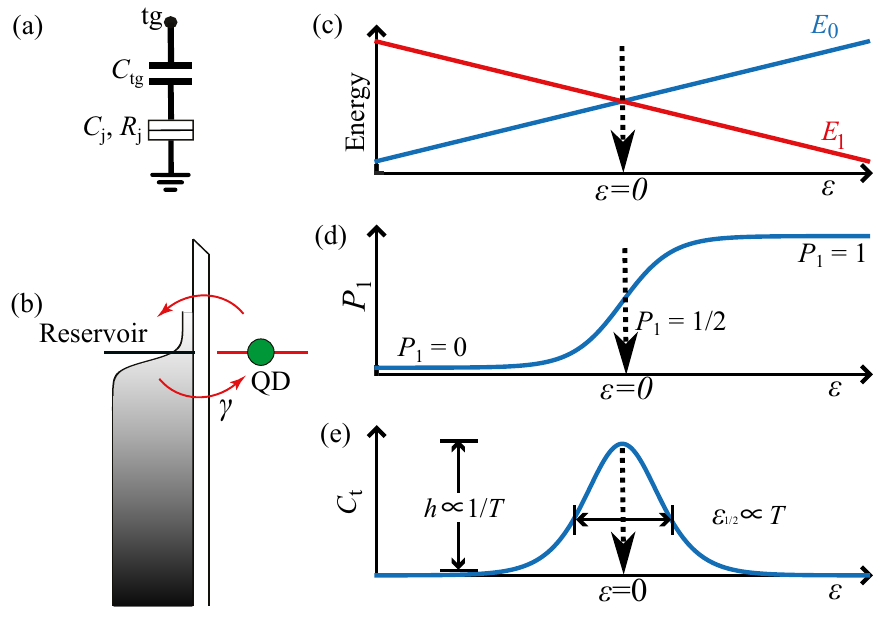}
\end{center}
\caption{ (a) Circuit equivalent of the QD-reservoir system. (b) Schematic of cyclic electron exchange between a discrete energy level of a QD and a thermally broadened electron reservoir. (c) Energy diagram of a fast driven TLS with discreet energies $E_0$ and $E_1$, across a charge degeneracy point. (d) Probability $P_\text{1}$ of an electron to be in the QD as a function of energy level detuning $\varepsilon$. (e) Tunneling capacitance $C_\text{t}$ as a function of $\varepsilon$.}
\label{fig1}
\end{figure} 

We consider a QD in thermal equilibrium with an electron bath whose temperature $T$ we wish to measure. The QD is capacitively coupled to a gate electrode $C_\text{tg}$, and tunnel coupled to the reservoir via a tunnel junction with capacitance $C_\text{j}$ and resistance $R_\text{j}$, see Fig.~\ref{fig1}(a). The system is operated in the quantum confinement regime such that electrons occupy discrete energy levels of the QD. The coupled QD-reservoir system has an associated differential capacitance~\cite{mizuta2017quantum, shevchenko2012multiphoton} as seen from the gate, $C_\text{diff}$, given by


\begin{equation}\label{diffCap}
	C_\text{diff}=\frac{\partial Q}{\partial V_\text{tg}}=\underbrace{\alpha C_\text{j}\rule[-15pt]{0pt}{1.7pt}}_{\mbox{\footnotesize geometrical}}-\underbrace{e\alpha\frac{\partial P_1}{\partial V_\text{tg}}\rule[-15pt]{0pt}{1.7pt}}_{\mbox{\footnotesize tunneling}},
\end{equation}

\noindent where $Q$ is the net charge in the QD, $V_\text{tg}$ is the gate voltage, $e$ is the electron charge, $\alpha$ is the gate coupling $C_\text{tg}/(C_\text{j}+C_\text{tg})$ and $P_1$ is the probability of having an excess electron in the QD. The first term in Eq.~\ref{diffCap} represents the DC limit of the capacitance, the geometrical capacitance, whereas the second term represents the parametric dependence of the excess electron probability on gate voltage, the tunneling capacitance. The second term is the focus of this Article. 

To obtain an analytical expression for the tunneling capacitance $C_\text{t}$, we next consider the QD-reservoir charge distribution in detail. In the limit of weak tunnel coupling, the QD-reservoir system can be described by the Hamiltonian $H=\frac{1}{2}\varepsilon\sigma_\text{z}$ where $\varepsilon$ is the energy detuning and $\sigma_\text{z}$ is the z Pauli matrix. The eigenergies $E_0=\varepsilon/2$ and $E_1=-\varepsilon/2$ are associated with the QD states with zero and one excess electron, respectively. This additional electron can tunnel in and out of the electron reservoir at a rate $\gamma$, as schematically depicted in Fig.~\ref{fig1}(b). The energy detuning between these states can be controlled by $V_\text{tg}$ given that $\varepsilon=-e\alpha(V_\text{tg}-V_\text{0})$. Here $V_\text{0}$ is the gate voltage offset at which the two eigenstates are degenerate. 

To probe the tunneling capacitance, the system is subject to a modulation occurring at some frequency, $f_\text{r}$ that varies the energy detuning $\varepsilon=\varepsilon_0+\delta\varepsilon\sin(2\pi f_\text{r}t)$. In the limit $\gamma>f_\text{r}$, the QD and reservoir are in thermal equilibrium and electrons tunnel in and out of the reservoir adiabatically. In this situation, $P_1$ tracks the thermal population given by the instantaneous gate-voltage excitation~\cite{mizuta2017quantum} and $C_\text{t}$ can be expressed as

\begin{equation}
C_\text{t}=-e\alpha\frac{\partial P_1^0}{\partial V_\text{tg}}=(e\alpha)^2\frac{\partial P_1^0}{\partial\varepsilon}.
\end{equation}

From the energy spectrum represented in Fig.~\ref{fig1}(c), Maxwell-Boltzmann statistics give the equilibrium probability distribution

\begin{equation}\label{prob}
P_1^0 = \frac{\exp(\varepsilon/2k_\text{B}T)}{\exp(-\varepsilon/2k_\text{B}T) + \exp(\varepsilon/2k_\text{B}T)},
\end{equation}

\noindent and this is depicted as a function of detuning in Fig.~\ref{fig1}(d). At large negative detuning the QD remains unoccupied ($P_1^0=0$), at large positive detuning the QD is occupied ($P_1^0=1$) and at the degeneracy point $P_1^0=1/2$. We calculate the tunneling capacitance of the system and obtain

\begin{equation}\label{tunnel}
C_\text{t} = \frac{(e \alpha)^2}{4k_\text{B}T} \frac{1}{\cosh^2(\frac{\varepsilon}{2k_\text{B}T})}.
\end{equation}

Thus, $C_\text{t}$ has a full width at half maximum (FWHM) with respect to $\varepsilon$ of

\begin{equation}\label{fwhm}
\varepsilon_{1/2}=4\ln(\sqrt{2}+1)k_\text{B}T, 
\end{equation}

\noindent as plotted in Fig.~\ref{fig1}(e). Since $\varepsilon_{1/2}=e\alpha V_{1/2}$, the analysis shows it is possible to obtain the temperature of the electron reservoir from the FWHM of the $C_\text{t}$ vs $V_\text{tg}$ curve once the gate lever arm $\alpha$ is known. $V_\text{1/2}$ is the FWHM with respect to gate voltage. Furthermore, we see that the absolute value of the tunneling capacitance $C_\text{t}^0$, is inversely proportional to the reservoir temperature,

\begin{equation}\label{hight}
C_\text{t}^\text{0} \propto \frac{1}{T}.
\end{equation}

We note that our analysis is valid as long as $k_\text{B}T$ remains smaller than the discrete energy spacing in the QD ($\Delta E$) and larger than the QD level broadening ($h\gamma$). These two conditions set the temperature range in which thermometry by cyclic electron tunneling is accurate. In the latter case ($k_\text{B}T<h\gamma$), $C_\text{t}$ takes a Lorentzian form given by 

\begin{equation}\label{lowT}
C_\text{t}=\frac{(\alpha e)^2}{\pi}\frac{h\gamma}{(h\gamma)^2 +\varepsilon^2}.
\end{equation}

\noindent and $\varepsilon_{1/2}$ is given by $2h\gamma$~\cite{cottet2011mesoscopic}, and is thus no longer temperature dependent. The relaxation rate $\gamma$ is directly linked to the shape of the tunnel barrier between the QD and the reservoir which can be tuned electrically by, for example, a gate electrode. $C_\text{t}$ can be probed with high-frequency techniques such as gate-based reflectometry~\cite{Colless2013,gonzalez2015probing} and can be used to measure temperature. We refer to this sensor as the gate-based electron thermometer (GET).

\section{\label{sec:level2} Device and high-frequency resonator}

The device used here is a silicon nanowire field-effect transistor (NWFET) \cite{betz2014high} fabricated in fully-depleted silicon-on-insulator (SOI) following CMOS rules. At low temperatures, gate-defined QDs form in the channel of the NWFET~\cite{Voisin2014,Anasua2018}, see Fig. 2(a). The transistor has a channel length $l=44$~nm and width $w=42$~nm. The 8 nm thick NW channel was pattered on SOI above the 145 nm buried oxide (BOX). The gate oxide consists of 0.8~nm SiO$_2$ and 1.9~nm HfSiON resulting in an equivalent gate oxide thickness of 1.3 nm. The top-gate (tg) is formed using 5~nm TiN and 50~nm polycrystalline silicon. The NW channel is separated from the highly doped source and drain reservoirs by 20 nm long Si$_3$N$_4$ spacers. The silicon wafer under the BOX can be used as a global back-gate (bg). 

\begin{figure}[H]
\begin{center}
\includegraphics[scale=1]{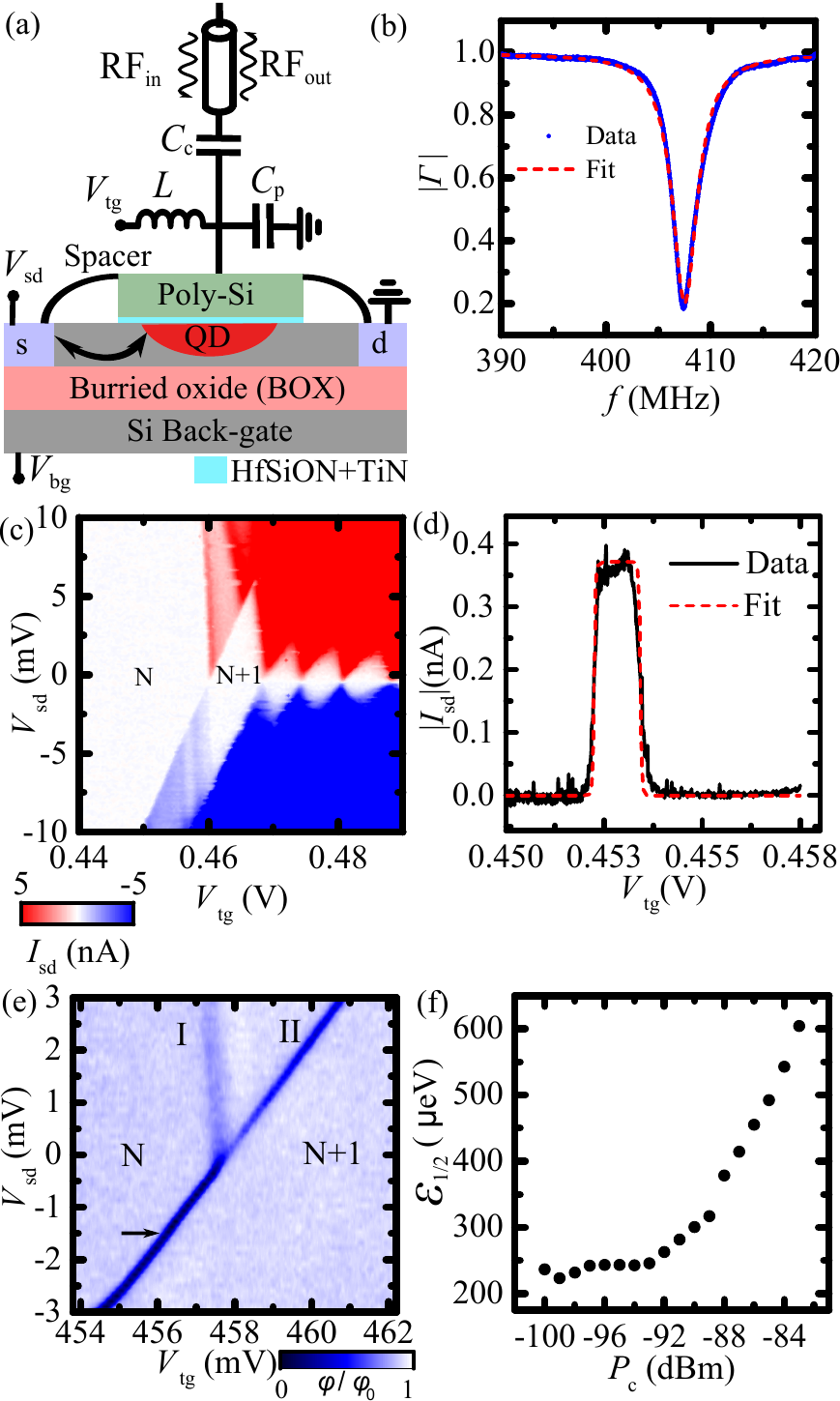}
\end{center}
\caption{ (a) Schematic of the device along the NW showing the source (s), drain (d), top-gate (tg) and back-gate (bg) terminals. The LC resonator is formed with a surface mount inductor $L$ connected to the top-gate and the parasitic capacitance to ground $C_\text{p}$. $C_\text{c}$ decouples the resonator from the line. $V_\text{sd}$, $V_\text{tg}$ and $V_\text{bg}$ are source, top-gate and back-gate bias voltages. (b) The amplitude $\left|\mathit{\Gamma}\right|$ of the reflection coefficient as function of frequency: data in blue and a Lorentzian fit in red. (c) Source to drain current $I_\text{sd}$ as a function of $V_\text{sd}$ and $V_\text{tg}$ showing Coulomb diamonds. (d) $I_\text{sd}$ trace as as a function of $V_\text{tg}$ at $V_\text{sd}$= -1.5 mV: data in black and fit in red. The arrow indicates the bias region for thermometry. (e) Relative demodulated phase $\phi/\phi_0$ as a function of $V_\text{sd}$ and $V_\text{tg}$ showing the stability map of the first electronic transition. The symbol I(II) indicates the electronic transition from source(drain) to QD. $N$ and $N+1$ indicate the bias regions with fixed electron number. (f) $\varepsilon_{1/2}$ of the charge transition line II as function of rf-carrier power $P_\text{c}.$} 
\label{fig2}
\end{figure}

To probe the device tunneling capacitance, we embed the transistor in a resonator formed by a 470 nH inductor -- connected to the top-gate (tg) of the device --- and the device parasitic capacitance $C_\text{p}$, which appears in parallel with the differential capacitance of the device, as can be seen in Fig. 2(a). We couple the resonator to a high-frequency line via a coupling capacitor $C_\text{c}=130~$fF. In order to characterize the resonator, we measure the reflection coefficient $\mathit{\Gamma}$. In Fig.~\ref{fig2}(b), we plot $\left|\mathit{\Gamma}\right|$ (data in blue and a fit in red) as a function of frequency $f$ at a fixed back-gate voltage $V_{\text{bg}}=3~$V. We extract the resonator's natural frequency of oscillation, $f_\text{0} = 1/(2\pi\sqrt{L(C_\text{c}+C_{\text{p}})})=408$~MHz, the bandwidth BW=2.9~MHz, the loaded quality factor $Q_{\text{L}}=141$ and $C_\text{p}=194$~fF. We find that the resonator is overcoupled but the depth of resonance, $\left|\mathit{\Gamma}\right|_\text{min}=0.18$ indicates that the resonator is close to being matched to the line.

\section{\label{sec:level2} The nature of cyclic electron tunneling}

A system with discrete energy levels $E_\text{0}$ and $E_\text{1}$ as described in Section II, can be found in a 0D QDs where the DOS consists of a series of delta functions at discrete energies~\cite{harrison2016quantum}. In this section, we attempt to demonstrate the discrete nature of the QD in NWFET using electrical transport measurements. 

We measure the source-drain current $I_\text{sd}$ as function of $V_\text{tg}$ and source-drain voltage $V_\text{sd}$. $I_\text{sd}$ shows characteristic Coulomb blockade diamonds when measured as a function of $V_\text{tg}$ and $V_\text{sd}$, see Fig. 2(c). Coulomb blockade diamonds are a signature of sequential single-electron transport through the QD from the source (s) to drain (d) reservoir. From the height of the Coulomb diamond in the charge stable configuration, we extract the QD first addition energy, $E_\text{add}=6$~meV, and the charging energy $E_\text{C}=3.75$~meV, indicating that $\Delta E=2.25$~meV.



When the QD has a 0D DOS and the source(drain) reservoirs have a 3D DOS, then Fermi's golden rule yields for the source(drain) tunnel rate 

\begin{equation}\label{tunel}
\gamma_{\text{s(d)}} = \frac{\gamma_\text{0,s(d)}}{1 + \exp(-\varepsilon_\text{s(d)}/k_\text{B}T)},
\end{equation}

\noindent where $\varepsilon_\text{s(d)}$ is the level detuning between the QD and s(d) reservoirs and $\gamma_\text{0,s(d)}$ is the tunnel rate at $\varepsilon_\text{s(d)}=0$~\cite{gonzalez2015probing}. Note that these tunnel rates are significantly different from metallic (3D DOS) QDs tunnel coupled to 3D reservoirs~\cite{persson2010excess}. Assuming that a single discrete energy level of the QD is within the energy window $eV_\text{sd}$, the source drain current $I_\text{sd}$ can be written in terms of tunneling rates $\gamma_\text{s}$ and $\gamma_\text{d}$ by the relation $I_\text{sd}=e\gamma_\text{s}\gamma_\text{d}/(\gamma_\text{s}+\gamma_\text{d})$~\cite{Beenakker1991} and is fitted to the data measured at fixed $V_\text{sd}=-1.5$~mV  in Fig.~\ref{fig2}(c). The agreement between the data and the fit demonstrates the 0D nature of the QD, showing it is suitable for the electron thermometry method introduced in Section II. 



\section{\label{sec:level2} Gate coupling and optimal power}

In order to get an accurate reading of the temperature $T$ from Eq.~\ref{fwhm}, the gate lever arm $\alpha$ needs to be obtained. We use gate-based reflectometry techniques to probe the charge stability map of the QD in the voltage region of interest, see Fig.~\ref{fig3}(e). We excite the resonator at resonant frequency $f_\text{0}$ and monitor the reflected signal. We used standard homodyne detection techniques~\cite{gonzalez2015probing} to measure the demodulated phase response $\varphi$ of the resonator as a function of $V_{\text{sd}}$ and $V_{\text{tg}}$. The phase of the resonators changes (dark blue lines I and II in Fig.~2(e)) at the charge degeneracy points due to a tunneling capacitance contribution. The separation in $V_\text{tg}$ between I and II at a fixed $V_\text{sd}$ gives a measurement of $\alpha=0.90\pm 0.01$. This large value --close to 1-- is consistent with the multi-gate geometry and the small equivalent gate oxide thickness of 1.3~nm of NWFETs~\cite{gonzalez2015probing}. 



Finally, we calibrate the optimal power on the resonator using transition II at $V_\text{sd}=-1.5$~mV, which we will subsequently use to perform thermometry. In Fig.~\ref{fig2}(f), we plot $\varepsilon_{1/2}$ as a function of the carrier power $P_\text{c}$ at the input of the resonator. At high carrier power, $P_\text{c}>-93$~dBm, $\varepsilon_{1/2}$ increases with $P_\text{c}$ indicating the transition is power broadened. For $P_\text{c}<-93$~dBm, $\varepsilon_{1/2}$ remains independent of $P_\text{c}$ and hence, we observe the intrinsic linewidth of the transition. We select $P_\text{c}=-95$~dBm hereinafter.  


\section{\label{sec:level2} Primary thermometry}

\begin{figure*}
\begin{center}
\includegraphics[scale=1]{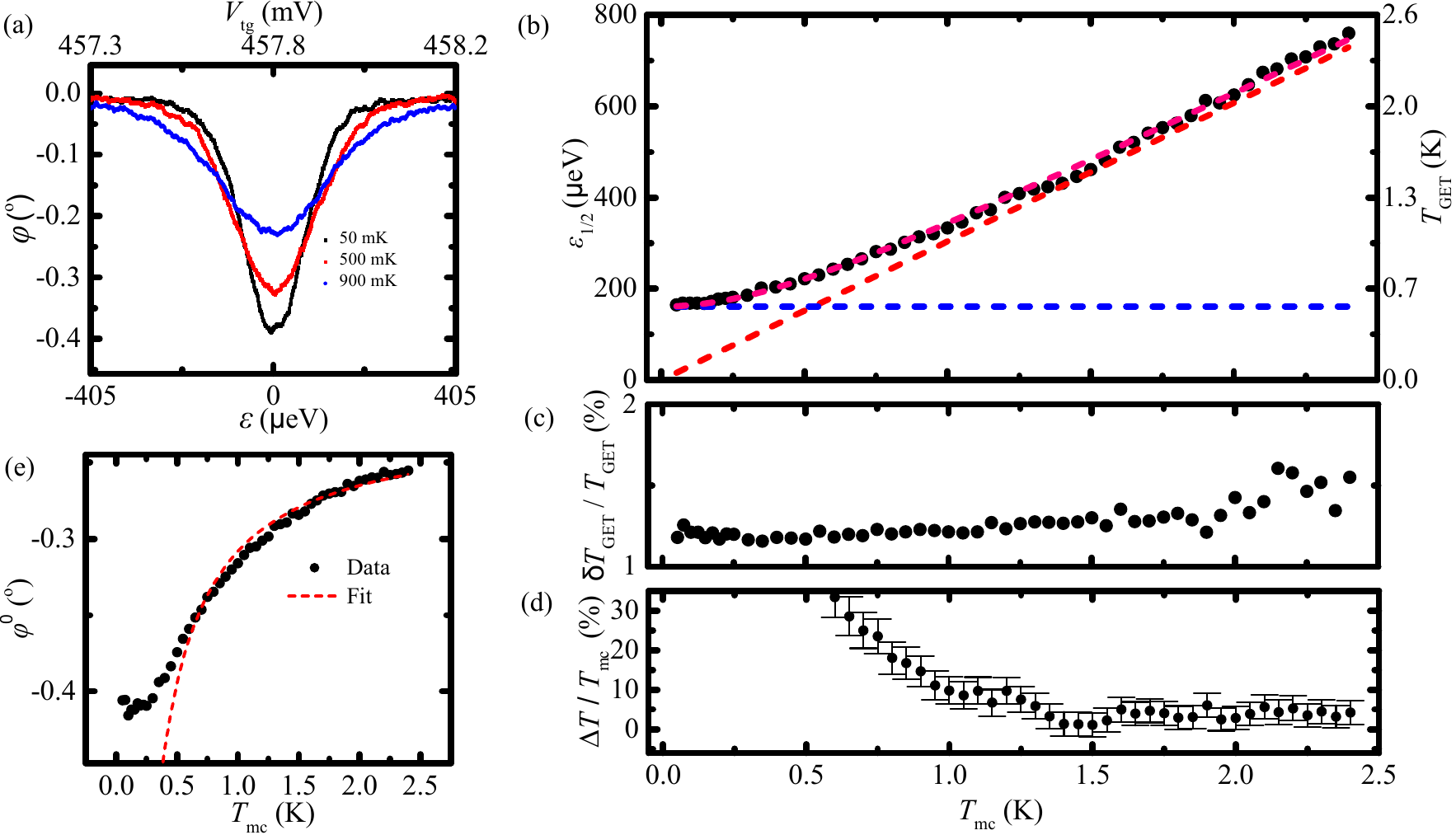}
\end{center}
\caption{ (a) Phase response $\varphi$ of the resonator as a function of top-gate voltage $V_\text{tg}$ swept across a charge degeneracy point for different $T_\text{mc}$. (b) $\varepsilon_\text{1/2}$ and $T_\text{GET}$ as a function of mixing chamber temperature $T_\text{mc}$ (black dots). Theoretical predictions to the low temperature regime (dashed blue), high temperature regime (dashed red) and full temperature range (dashed magenta). (c) Fractional temperature precision $\delta T_\text{GET}/T_\text{GET}$ and (d) fractional accuracy $\Delta T/T_\text{mc}$ as a function of $T_\text{mc}$. $\Delta T=T_\text{GET}-T_\text{mc}$. (e) Phase change at $\varphi^0$ as a function $T_\text{mc}$ (black dots) and a $1/T_\text{mc}$ fit at high temperature $T_\text{mc}>1K$ (red dashed line).}
\label{fig3}
\end{figure*}

In this section, we explore experimentally gate-based primary thermometry using transition II (see Fig.~\ref{fig2}(f)). As we have seen in Section II, when $k_\text{B}T/h>\gamma>f_0$, electron tunneling between QD and reservoir has an associated tunneling capacitance whose $\varepsilon_{1/2}$ gives a reading of the reservoir temperature (see Eq.~\ref{fwhm}). In this experiment, we probe $T$ from a measurement of $\varphi$ vs $\varepsilon$, since $\varphi=-2Q_\text{L}C_\text{t}/C_\text{p}$~\cite{chorley2012measuring,gonzalez2016gate,betz2015dispersively}, when the resonator is overcoupled to the line.  We drive the resonator at frequency $f_\text{0}$ and monitor $\varphi$ as we sweep $\varepsilon$ across the charge degeneracy for different temperatures of the mixing chamber $T_\text{mc}$, see Fig. 3(a). We measure $T_\text{mc}$ with a 2200~$\Omega$ RuO$_2$ resistive thermometer. As the temperature is increased, $\varepsilon_{1/2}$ increases and the maximum phase shift decreases. We repeat the measurement for several $T_\text{mc}$ and plot $\varepsilon_{1/2}$ in Fig.~\ref{fig3}(b). Two clear temperature regimes become apparent:

At low temperatures, for $T_\text{mc}<200$~mK, we see that $\varepsilon_{1/2}$ is independent of $T_\text{mc}$ and equal to $160~\mu$eV (blue dotted line). In this regime, as we shall demonstrate later, the thermal energy is smaller than the QD level broadening ($\text{k}_\text{B}T<h\gamma$). As a result, the temperature reading of the GET, $T_\text{GET}$, deviates from the mixing chamber thermometer. On the other hand, at high temperatures, $T_\text{mc}>1$~K, we observe that $\varepsilon_{1/2}$ presents a linear dependence with $T_\text{mc}$ as predicted by Eq.~\ref{fwhm}. For comparison, we plot the theoretical prediction (red dashed line) and observe that both follow a similar trend. In this regime, since $h\gamma<\text{k}_\text{B}T$, the GET can be used to obtain an accurate reading of the temperature of the electron reservoir. We quantify the precision of the thermometer by measuring the fractional uncertainty in the temperature reading of the gate-based thermometer, $\delta T_\text{GET}/T_\text{GET}$ (see Fig.~\ref{fig3}(c)). At low temperatures, the precision of the thermometer is primarily determined by the uncertainty in the lever arm, $\delta\alpha/\alpha=1.1\%$. As we raise the temperature, the phase response of the resonator becomes smaller leading to an increase in the uncertainty of $V_{1/2}$ which, at the highest temperatures, becomes comparable to that of $\alpha$. We find $\delta T_\text{GET}/T_\text{GET}$ increases up to $1.6\%$. Additionally, in Fig.~\ref{fig3}(d), we determine the fractional accuracy of the GET thermometer, $\Delta T/T_\text{mc}$ by comparing its reading with that of the RuO$_2$ thermometer ($\Delta T=T_\text{GET}-T_\text{mc}$). We see than the discrepancy between thermometers is less than 8\% for temperatures higher than 1~K and this goes down to an average of 3.5\% above 1.5~K. The error in the accuracy is primarily determined by the uncertainty in the reading of the RuO$_2$ thermometer, which varies from 1\% at the lowest temperatures to 6\% at 2.4~K, rather than by the precision of the GET.

We note that, although not applicable for primary thermometry purposes, the whole temperature range can be described by a single expression that combines both regimes, level-broadening and thermal broadening, in to a single expression $\varepsilon_{1/2}=\sqrt{(3.53k_\text{B}T)^2+(2h\gamma)^2}$ (see purple dashed line). This formula fits well the data and we find that the difference is less than $6\%$ for all temperatures. 


Lastly in Fig.~\ref{fig3}(e), we measure the phase shift at the degeneracy point $\varphi^0$, as a function of $T_\text{mc}$. Again, the two regimes are apparent. At low temperatures $\varphi^0$ remains constant and only at temperatures $T_\text{mc}>1$~K, $\varphi^0$ shows an inverse proportionality with $T_\text{mc}$ as predicted by Eq.~\ref{hight} (dashed red line).

\section{\label{sec:level2} Low Temperature Limit}



In Fig.~\ref{fig3}(b,e), we have seen that at low temperatures both $\varepsilon_{1/2}$ and $\varphi^0$ deviate from the prediction in Sec. II. In this regime, the gate-sensor cannot be used as an accurate thermometer. Two mechanisms may be responsible for this discrepancy: Electron-phonon decoupling, due to the weaker interaction at low $T$~\cite{maradan2014gaas,bradley2016nanoelectronic}, or lifetime broadening, when the QD energy levels are broadened beyond the thermal broadening of the reservoir, which occurs when $h\gamma>k_\text{B}T$. In the latter case, $\varepsilon_{1/2}$ is given by $2h\gamma$ (see Eq.~\ref{lowT}) whereas for the former, it is given by $3.53k_\text{B}T_\text{dec}$, where $T_\text{dec}$ is the decoupling temperature. 

\begin{figure}
\begin{center}
\includegraphics[scale=1]{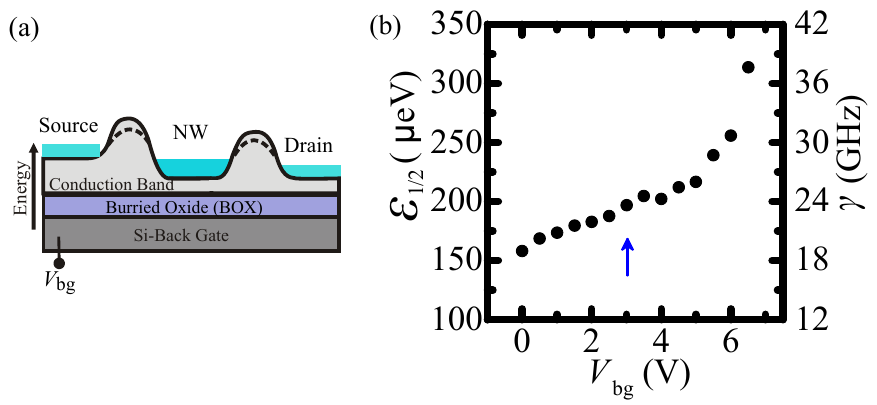}
\end{center}
\caption{(a) Schematic of the conduction band edge along the NW channel. The QD-reservoir tunnel barriers can be controlled \lq\lq in-situ\rq\rq. (b) $\varepsilon_{1/2}$ and $\gamma$ as a function of back-gate voltage $V_\text{bg}$. The arrow indicates the $V_\text{bg}$ at which thermometry was performed.}
\label{fig4}
\end{figure}

To assess the origin of the discrepancy, we modify the tunnel barrier potential by varying the vertical electric field across the device (Fig.~\ref{fig4}(a)) which effectively changes $\gamma$~\cite{gonzalez2015probing}. We do so by changing the potential on the back-gate electrode $V_\text{bg}$ while compensating with $V_\text{tg}$. In Fig.~\ref{fig4}(b), we plot $\varepsilon_{1/2}$ as a function of $V_\text{bg}$. We see that as we lower $V_\text{bg}$, $\varepsilon_{1/2}$ decreases, indicating that the tunnel rate $\gamma$ across the potential barrier is lower due to the increasing height of the potential barrier at lower $V_\text{bg}$. This trend indicates that at low temperature, our primary thermometer is limited by level broadening and not by electron-phonon decoupling. Moreover, it demonstrates it is possible to tune electrically the low temperature range of the primary thermometer, as long as $\gamma$ remains larger than the excitation frequency $f_\text{0}$.


\section{\label{sec:level2} Conclusion}

We have described and demonstrated a novel primary electron thermometer based on cyclic electron tunneling that allows measuring the temperature of a single electron reservoir without the need of electrical transport. The GET requires of a system with at least one discrete energy level tunnel-coupled to the reservoir to be measured an scenario that can be found in single-molecule junctions and/or in single-electron devices. Here, we have implemented the thermometer using CMOS technology which makes it ideal for large-scale production. Conceptually, the GET is simpler than the CBT as it requires just a single tunnel junction for operation. Driving and readout of the thermometer can be performed simultaneously using reflectometry techniques which have recently demonstrated high-sensitivity with MHz bandwidth~\cite{ahmed2018radio}. This feature makes it potentially faster than more resistive thermometers such the SIN-junction thermometer or the CBT and of comparable acquisition rate than the SNT. However, the GET is unlikely to have the large dynamic range of the latter since, the high temperature limit is set by quantum confinement. To achieve room temperature operation, in the case of a silicon-based GET, it would require sub-5~nm device dimensions and sensitive capacitance meters down to the attoFarad range. Nevertheless, due to the adiabatic nature of electron tunneling, the GET is likely to show lower self-heating that the aforementioned thermometers. We have shown accurate primary thermometry down to 1~K and have proven that the low temperature range can be electrically tuned \lq\lq in-situ\rq\rq~. For sub-10~mK operation, low transparency barriers and driving resonators with sub-200~MHz resonant frequencies should be used to ensure the thermometer is operated in the adiabatic limit. Overall, our thermometer shows potential for local probing of fast heat dynamics in nanoelectronic devices and it may have applications in the better study of thermal single-electron devices such as rectifiers and energy harvesters. Moreover, since the device is made using silicon technology it could naturally be integrated with silicon-based quantum circuits.

 %
%

\section{\label{sec:level2} Acknowledgments}

We thank Jonathan Prance for providing useful comments. This research has received funding from the European Union's Horizon 2020 Research and Innovation Programme under grant agreement No 688539 (http://mos-quito.eu) and the Winton Programme of the Physics of Sustainability. IA is supported by the Cambridge Trust and the Islamic Development Bank. AC acknowledges support from the EPSRC Doctoral Prize Fellowship.


\bibliographystyle{apsrev4-1}
\bibliography{Thermo}
\end{document}